\documentclass [12pt]{umj}
\usepackage[cp1251]{inputenc}
\usepackage[english]{babel}
\usepackage{amsmath}
\usepackage{amssymb}
\usepackage{amsfonts}
\def\udcs{517.9} 

\firstpage{127}
\subjclass{\udcs}

\renewcommand{\No}{No}

\begin{document}
\thispagestyle{empty}

\title[``Quantum'' linearization  of Painlev\'{e} equations \ldots]{``Quantum'' linearization  of Painlev\'{e} equations as a component of their $L,A$ pairs}

\author{B.I. Suleimanov}

\address{Bulat Irekovich Suleimanov,
\newline\hphantom{iii} Institute of Mathematics with Computer Center of Ufa Science Center of the Russian Academy of Sciences,
\newline\hphantom{iii} 112, Chernyshevsky str., 
\newline\hphantom{iii} Ufa, Russia, 450008}
\email{bisul@mail.ru}

\thanks{\sc B.I. Suleimanov,  
``Quantum'' linearization  of  Painlev\'{e} equations as a component of their $L,A$ pairs.}%
\thanks{\copyright \ Suleimanov B.I. 2012}
\thanks{\rm The paper is supported by FTP ( contract 02.740.11.0612 ).
}
\thanks{\it Submitted on March 1, 2012.}

\maketitle {\small
 \begin{quote}
\noindent{\bf Abstract.} The procedure  of the ``quantum'' linearization of the Hamiltonian ordinary differential equations with one degree of freedom is 
introduced. It is offered to be used for the classification of  integrable equations of the Painleve type.
By this procedure and all natural numbers $n$ we construct the solutions $\Psi(\hbar,t,x,n)$ to the non-stationary   Shr\"{o}dinger equation with the Hamiltonian $H = (p^2+q^2)/2$ 
which tend to zero as $x\to\pm\infty$. On the curves $x=q_n (\hbar, t) $ defined by the old Bohr-Sommerfeld quantization rule the solutions satisfy the relation  
$i\hbar \Psi ' _x\equiv p_n (\hbar, t) \Psi $, where   $p_n (\hbar, t) = (q_n (\hbar, t)) ' _t $
is the classical momentum corresponding to the harmonic $q_n (\hbar, t) $.
\medskip

\end{quote}
}

\section{Introduction}

Among the procedures applied to nonlinear equations  linearization is perhaps the most widely spread one. It is applied not only for working with the approximations to solutions.

In particular, experts in equations integrated by the method of inverse scattering transform know that these equations possess true ``working'' $L$,$A$ pairs, one component of which is the result of linearization. For example, for the Korteweg--de Vries equation $$u'_t=u'''_{xxx}+uu'_x$$ this component is of the form $$U'_{t}=U'''_{xxx}+uU'_x+u'_xU.$$ The second component of such pairs is determined by the 
Hamiltonian structures of equations accepting this method 
(by  the recursion operator \cite{Olv}).

{\it Remark $1$.} It is not excluded that all such pairs are equivalent to the usual ones. Anyway, considering the case of $L,A$ pairs of the sin-
Gorgon equation $$u''_{xt}+\sin u=0$$ hints on this assumption. Its traditionally employed pair consists of \cite{Abl}  the system of linear equations ($\zeta$ is a spectral parameter)
\begin{equation}\label{sinu}(v_1)'_x=-i\zeta v_1+Qv_2,\quad (v_2)'_x=-Qv_1+i\zeta v_2,\end{equation}
\begin{equation}\label{sinv}4i\zeta(v_1)'_t=\cos (u) v_1+\sin (u) v_2,\quad 4i\zeta 
(v_2)'_t=(\sin u)v_1-(\cos u) v_2,\end{equation}
where $Q=-u_x/2$. O. M. Kiselev \cite [\S4.1.1]{kisl} showed, that the combination of squares
$$\Phi^{\pm}(x,t,\zeta)=v_2^2(x,t,\zeta)\pm v_1^2(x,t,\zeta),\qquad \Psi(x,t,\zeta)=v_1(x,t,\zeta)v_2(x,t,\zeta)$$
of the solutions $L,A$ of the pair (\ref{sinu}), (\ref{sinv}) satisfies to two linear systems of ordinary differential equations (ODE)
\begin{equation}\label{lins}(\Phi^+)'_x=2i\zeta\Phi^-,\quad
(\Phi^-)'_x=2i\zeta\Phi^+-4Q\Psi,\quad\Psi'_x=Q\Phi^-,\end{equation}
\begin{equation}\label{linst}2i\zeta(\Phi^+)'_t=\sin (u)\Psi,\quad2i\zeta(\Phi^-)'_t=-\cos (u)\Phi^+,\quad 2i\zeta\Psi'_t=Q\sin(u)\Phi^+\end{equation}
and noticed that the component $\Phi^+$ of the solution of the system (\ref{lins}), (\ref{linst}) satisfies the equation $$(\Phi^+)_{xt}+\cos (u)\Phi^+=0,$$ being the  result of linearization of the sin-Gordon equation. This conclusion of O.M.~Kiselev is supplemented by the following observation:

- the component $\Phi^+$ of the solution of the system (\ref{lins}) {,(\ref{linst})}
 also satisfies the equation $$-4\zeta^2\Phi^+=(\Phi^+)''_{xx}+4Q^2\Phi^+-
4Q \int^x Q'_x\Phi^+dx,$$ the right side of which is the result of the influence on $\Phi^+$ of the recursion operator for the sin-Gordon equation written by A.V. Zhiber and A.B. Shabat in their 
well-known paper \cite{Zhi}.

Alongside with the result of linearisation after creating of the wave quantum mechanics one often associates to the Hamiltonian systems of ODE
\begin{equation}\label{Hams}
\lambda'_{\tau}=H'_{\mu}(\tau,\lambda,\mu),\quad \mu'_{\tau}=-H'_{\lambda}(\tau,\lambda,\mu),\end{equation}
whose Hamiltonians are
quadratic w.r.t. the impulse
$\mu$ 
\begin{equation}\label{hamv} H(\tau,\lambda,\mu)=\alpha(\tau,\lambda)\mu^2+\beta(\tau,\lambda)\mu +\gamma(\tau,\lambda),
\end{equation}
another linear differential equation which is the non-stationary Schr\"{o}dinger equation ($\hbar$ is the Planck constant)
\begin{equation}\label{Shro}
i\hbar\Psi'_{\tau}=H(\tau,z,-i\hbar\frac{\partial}{\partial z})\Psi.
\end{equation}
It happens \cite{Ufa88} --- \cite{Garfn} that for all Painleve ODE integrated by the method of isomonodromic deformations \cite{Ifnk}, the linear equations  similar to (\ref{Shro}) (like in \cite{Garfn}, in what follows they are called as ``quantizations'' of the second-order ODE on the coordinate $\lambda(\tau)$)
\begin{equation}\label{qquant}
\Psi'_{\tau}=H(\tau,z,\frac{\partial}{\partial z})\Psi \end{equation}
can be considered as a component of the corresponding $L,A$ pair.

{\it Remark $2$.} Such  ``quantization'' occur in the problems of filtration of diffusion processes \cite{Ovs}, \cite{Dovs}.

In the present paper we introduce the procedure associating 
Hamiltonian ODE with one more set of linear equations being certain ``quantum'' analogues of the results of 
linearization of the given ODE (particular examples of such ``quantum'' 
linearizations of some Painleve ODE are given below).
These linear equations and equations determined by the ``quantizations''  (\ref{qquant}) are supposed, in particular, to be employed as  $L,A$ pairs of the general kind for the classification of 
{Hamiltonian ODE (with the Hamiltonian $H$ of the kind (\ref{hamv}))} 
which can be integrated by the method of isomonodromic deformations.

The main part of the paper begins with the section devoted to a rather interesting aspect of the problem of quantization for the harmonic oscillator, which was found out exactly by such $L,A$ pair.

\section{
Trajectories of quasi-determinacy for the harmonic oscillator.}%

An old quantum theory with the help of the well-known
Bohr-Sommerfeld  
{quantization rule} \linebreak \cite[Ch.1,\S 15, formula (17)]{Mess} 
singled out a set of trajectories $q_n(t,\hbar)$ with the energies
\begin{equation}\label{borze} H=H(n)=n\hbar\quad (n=1,2,\dots,\infty).\end{equation}
among all the solutions of the equations of a harmonic oscillator
\begin{equation}\label{harmk}
q''_{tt}+q=0\end{equation}
with the Hamiltonian of energy
\begin{equation}\label{hamkh}
H(q,p)=\frac{1}{2}(p^2+q^2).
\end{equation}

Later, in contrast to this set, matrix and wave mechanics singled out another series of energies
\begin{equation}\label{heish} H=E(n)=(n+\frac{1}{2}) \hbar\quad (n=0,1,2,\dots,\infty)\end{equation}
among other possible energies (\ref{hamkh})
(first by Heisenberg \cite[formulae (22),(23)]{Hei}).
In particular, this series describes the eigenvalues
$E=E(n)$ of ODE
\begin{equation} -\frac{\hbar^2}{2}\Phi_{xx}''+\frac{x^2}{2}\Phi=E\Phi\label{stasg},
\end{equation}
with the associated eigenfunctions
\begin{equation} \Phi_n(x,\hbar)=\alpha_nH_n(\sqrt{\frac{1}{\hbar}}x) \exp(-\frac{1}{2\hbar}x^2), \label{permi}
\end{equation}
where $\alpha_n$ are constants and $$H_n(z)=(-1)^n\exp(z^2)(\frac{d^n}{dz^n}\exp(-z^2))\quad (n=0,1,\dots)$$ are Hermite polynomials\cite[Ch.12, \S7, \S8; Supplement B, section III]{Mess}.
Each $\Phi_n(x,\hbar)$  
defines a solution to the 
{Schr\"{o}dinger} equation
\begin{equation}
i\hbar \Psi_t'=
\frac{(-i\hbar)^2}{2}\Psi_{xx}''+\frac{x^2}{2} \Psi\label{nestag} 
\end{equation}
by the formula 
$$\Psi(t,x,\hbar)= \Phi(x,\hbar)\exp(-i\frac{E(n)t}{\hbar}).$$

But now we are intended to demonstrate that the series of harmonics $q_n(\hbar,t)$ determining the energies (\ref{borze}) is nevertheless singled out while considering the 
{Schr\"{o}dinger} equation  (\ref{nestag}), namely, for every natural $n$  we shall construct a smooth solution
$\Psi(\hbar,t,x,n)$ of this equation, which decays exponentially as $x\to \pm\infty$ 
 and {\it only as $x=q_n(\hbar,t)$ it satisfies the relation
$$i\hbar\Psi'_x(\hbar,t,x,n)\equiv p_n(\hbar,t)\Psi(\hbar,t,x,n),$$ where  $p_n(\hbar,t)=q_n(\hbar,t)'_t$ is 
the classical impulse corresponding to the coordinate $q_n(\hbar,t)$}. In other words,  the  action of the operator of the quantum-mechanical impulse 
on the function $\Psi(\hbar,t,x,n)$ differs from the result of multiplication $\Psi(\hbar,t,x,n)$ by the classical impulse only by a sign {(as $x=q_n(\hbar,t)$)}.
These solutions $\Psi(\hbar,t,x,n)$ are defined below by the formula (\ref{genshgar}), in which the function $Q_-(\hbar,x)$ coincides with the right hand
side of the identity (\ref{permi}), and the constants $c_1$ and $c_2$ which determine the corresponding harmonics $q_n(t,\hbar)$ by the  by the formula (\ref{genga}) are  
complex conjugated and their moduli are equal to $\sqrt{ n\hbar/2}$. This fact is deduced from the validity of a more general statement:

for the solution of the ODE (\ref{harmk}) ($c_1,c_2$ are arbitrary constants)
\begin{equation}\label{genga}q(t)=c_1\exp{(i t)}+c_2\exp{(-it)},
\end{equation}
to which by the impulse $p(t)=q'(t)$ and the value $E=2c_1c_2$
the Hamiltonian (\ref{hamkh}) are associated, 
each solution $Q_-(x,\hbar)$ of the ODE (\ref{stasg}) 
by the following result of the action of the creation and annihilation operators
$$(\hbar\frac{\partial}{\partial x}+x)Q_-=-2c_2Q_+,\qquad
(\hbar\frac{\partial}{\partial x}-x)Q_+=2c_1Q_-,$$
{gives} the solution
\begin{equation}\label{genshgar}\Psi= \exp(-\frac{iEt}{\hbar})(\exp(\frac{i t}{2})Q_++exp(-\frac{ it}{2})Q_-)\end{equation}
of the equation (\ref{nestag}) satisfying the identity
$$(i\hbar\frac{\partial}{\partial x}-q'(t))\Psi=2(x-q(t))(\Psi'_t+
\frac{iE}{\hbar} \Psi).$$

{\it Remark $3$.} It was noted in the abridged paper \cite{mgu11} that the separation of classical trajectories $q(t)$ corresponding to the 
{old} version of the Bohr-Sommerfeld quantization rule is also observed for the discrete series of the solutions of the Schr\"{o}dinger equations (\ref{Shro}) determined by the $L, A$ pairs and Hamiltonians $H(q,p)$ of the series of autonomous reductions for the third and the fifth Painleve equations.  The author is planning to devote a separate paper to the description of the properties of such series of the solutions to the equations (\ref{Shro}) (corresponding to the well-known Morse and 
P\"{o}schl-Teller potentials) . In this paper we just note that the choice of a Hamiltonian $H(q,p)$ is important.  The importance of such a choice is clear, for instance, from comparison of the results \cite{Ufa88}, \cite{Difufa} with the results \cite{Zabr} for the fourth, the fifth, and the sixth Painleve equations. (see \cite{Zig}, \cite{Zigt} for the variety of the Hamiltonian structures of Painlev\'e equations.)

{\it Remark $4$.} 
For arbitrary values of $\varepsilon$ H. Nagoya  \cite{Nag}
constructed explicit (in terms of hyperheometric functions) solutions to 
the equations $\varepsilon\frac{\partial }{\partial t}\Psi=
H(t,x,\varepsilon\frac{\partial}{\partial x})\Psi,$
corresponding to the Hamiltonians of the series of reductions for   Painlev\'e equations. In the author's opinion, it is worth to study the question on which of these solutions as $\varepsilon=i\hbar$  are singled out by the boundedness  for all real $x$.

{\it Remark $5$.}  For each classical trajectory of the harmonic oscillator $g(\hbar,t)$ it is known \cite{Bab}, \cite{Babd} 
the solution to Schr\"{o}dinger equation (\ref{nestag})   which tends to zero as $x\to\pm \infty$ (this solution is concentrated in an exponentially small as $\hbar<<1$ neighbourhood of the curve $x=q(\hbar,t)$ ), and which satisfies the identity $$-i\hbar\Psi'_x(\hbar,t,x,n)\equiv q(\hbar,t)\Psi(\hbar,t,x,n)$$ as $x=q(\hbar,t)$.
But these solutions do not single out any discrete set of classical trajectories.

\section {``Quantization'' and ``quantum''
 linearization of Painlev\'e type equations}

The author came to the statement formulated above Remark 3 after his consideration of ``quantum'' nature of $L,A$ pair for ODE
($a_j$ are arbitrary constants)
 \begin{equation}\label{peodg}\lambda_{\tau \tau}''=a_4(2\lambda^3+\tau\lambda)+a_3(6\lambda^2+\tau)+a_2\lambda+a_1,\end{equation}
 stated by Garnier \cite[p.49]{Garn}.
This pair for the ODE (\ref{peodg}), which in particular cases contains the first and the second Painleve equations and also ODE, which is equivalent to (\ref{harmk}), 
is of  the form
\begin{equation} \label{garlpeodg}W_{zz}''=P(\tau,z)W,\qquad W_{\tau}'=B(\tau,z)W_z'-\frac{1}{2}B(\tau,z)'_zW,\end{equation}
where
$B=1/(2(z-\lambda)),$
$$P=a_4[z^4-\lambda^4+\tau(z^2-\lambda^2)]+2a_3[2(z^3-\lambda^3)+\tau(z-\lambda)]+$$
$$+a_2(z^2-\lambda^2)+2a_1(z-\lambda)+(\lambda')^2-
\frac{\lambda'}{z-\lambda}+\frac{3}{4(z-\lambda)^2}.$$
The equations (\ref{garlpeodg}) with such coefficients $B(\tau,z)$ and $P(\tau,z)$ are 
compatible for the solutions $\lambda(\tau)$ of ODE (\ref{peodg}).

The ``quantum'' nature of the present pair expresses, in particular, in the change
$$V=\sqrt{(z-\lambda)}W$$ 
which transforms the system (\ref{garlpeodg}) into the equations
\begin{equation} \label{rashpeod}V_{zz}''=\frac{V'_z}{z-\lambda}
{+}[P-\frac{3}{4(z-\lambda)^2}]V,\qquad
 V'_{\tau}=\frac{V_z-\lambda'V}{2(z-\lambda)},\end{equation}
whose simultaneous solution $V(\tau,z)$, as one easily see, satisfies the identity
\begin{equation}\label{kkpeod}V'_{\tau}=\frac{V_{zz}''}{2}-[\frac{a_4}{2}(z^4+\tau z^2)+a_3(2z^3+\tau z)+
\frac{a_2}{2}z^2+a_1z+H(\tau,\lambda(\tau),\lambda'(\tau)]V.
\end{equation}
Here the function $H(\tau,\lambda(\tau),\lambda'(\tau))$
as $\lambda=\lambda(\tau)$ and
$\mu=\lambda'(\tau)$ coincides with the Hamiltonian
$$H=\frac{\mu^2}{2}-\frac{a_4}{2}(\lambda^4+\tau \lambda^2)-a_3(2\lambda^3+\tau \lambda)-\frac{a_2}{2}\lambda^2-a_1\lambda$$
of the system (\ref{Hams}),
which is equivalent to the ODE (\ref{peodg}). The transformation {\linebreak}
$\Psi=\exp{(\int_{\tau_*}^{\tau}H(\nu,\lambda(\nu),\mu(\nu))d\nu)}V $
reduces the identity (\ref{kkpeod}) to the ``quantization'' of ODE (\ref{peodg})
$$\Psi'_{\tau}=\frac{\Psi_{zz}''}{2}-[\frac{a_4}{2}(z^4+\tau z^2)+a_3(2z^3+\tau z)+\frac{a_2}{2}z^2+a_1z]\Psi,$$
which does not explicitly depend on $\lambda(\tau)$.

In \cite{Ufa88}, \cite{Difufa} and for each of the other four canonic
Painlev\'e equations they found the Hamiltonian
$H=H_{j}(t,\lambda,\mu)\quad(j=3,\dots,6)$ determining
the Hamiltonian system (\ref{Hams}) equivalent to the corresponding
Painlev\'e ODE, and which is so that the ``quantization'' (\ref{qquant})
has the solutions 
set in fact by $L$,$A$ pairs from \cite{Garn}. (For the
third and the fifth Painlev\'e equations in the formulae \cite{Ufa88}, \cite{Difufa}
there are errors, however, which are easily corrected.)
Therefore the conclusion in the paper \cite{Garfn} on using the ``quantizations'' (\ref{qquant})  for the classification of Hamiltonian second-order ODE possessing $L,A$ pairs of the same type like Painlev\'e ODE looks logical. It is obvious that to make a classification, which could be as natural as the presented successful classifications of different classes of integrated equations  \cite{Zhi}, \cite{Sosv}---\cite{Habzp}, we need supplementary and in some sense natural restrictions.

Under the assumption that the first component of $L,A$ pairs of classified integrated Hamiltonian ODE is determined by the ``quantization'' (\ref{qquant}); below we suggest the general {\it ansatz} (\ref{genklin}) of their second component.  This {\it ansatz} generalizes, in particular, the form of ODE
\begin{equation}\label{klinpeod}4V''_{\tau\tau}=[ a_4(z^2+2z\lambda+3\lambda^2+\tau))+
4a_3(z+2\lambda)+a_2]V,\end{equation}
which, as it follows from the equations  (\ref{rashpeod}) and (\ref{kkpeod}), is simultaneously satisfied by their 
common solution.

ODE (\ref{klinpeod}) is the result of some ``quantum'' linearization of ODE (\ref{peodg});  by formal
``dequantization'' that is replacing in its right hand side $z$ by $\lambda(\tau)$ it transforms into the equation, which differs just by the multiplier 4  in its left side  from ODE $$\Lambda''_{\tau\tau}=[ (6\lambda^2(\tau)+\tau)a_4+12\lambda(\tau)a_3+a_2]\Lambda,$$ appearing as the result of the linearization of the ODE (\ref{peodg}).

As $a_4=a_3=a_1=0$  and $a_2=1$ ODE (\ref{peodg}) and the equation (\ref{klinpeod}) are reduced to the linear ODE with constant coefficients
\begin{equation}\label{trivg} 
\lambda''_{\tau\tau}=\lambda,
\end{equation}
\begin{equation}\label{klintr}
4V''_{\tau\tau}=V,
\end{equation}
and the equation (\ref{kkpeod}) is reduced to the linear equation
\begin{equation}\label{kkge}
V'_{\tau}=\frac{V_{zz}''}{2}-(\frac{z^2}{2}+H)V,
\end{equation}
where the Hamiltonian $H=(\mu^2(\tau)-\lambda^2(\tau))/2$ is constant.

It follows from the equations (\ref{klintr}) and (\ref{kkge}) that their simultaneous solution $V(\tau,z)$
is of the form
\begin{equation}\label{ansag}
V(\tau,z)=exp{(\frac{\tau}{2})}A_+(z)+\exp{(-\frac{\tau}{2})}A_-(z),
\end{equation}
where the functions $A_{\pm}(z)$ satisfy the linear ODE
\begin{equation}\label{vebpm}\frac{(A_{\pm})_{zz}''}{2}=(\frac{z^2}{2}+H \pm
\frac{1}{2})A_{\pm}.\end{equation}
Employing the validity of the second relation in (\ref{rashpeod}) for $V(\tau,z)$ makes it possible to specify that
for the solution of ODE (\ref{trivg}) ($r_1,r_2$ are arbitrary constants)
\begin{equation}\label{genl}\lambda(\tau)=r_1\exp{(\tau)}+r_2\exp{(-\tau)},
\end{equation}
and also the relations
\begin{equation}\label{rozh}
(A_+)'_z-zA_+=2r_1A_-,
\end{equation}
\begin{equation}\label{unich}(A_-)'_z+zA_-=-2r_2A_+
\end{equation}
hold true. These relations appear as a result of the substitution of the right hand side  (\ref{ansag}) into the result of the multiplication of the second equation in (\ref{rashpeod}) by the multiplier $z-\lambda(\tau)$ and the consequent comparison of the terms at various powers of $\exp({\tau})$.

For the solution  (\ref{genl})
$H=-2r_1r_2$. And it is easy to see that by any of the solutions $A_+$  $(A_-)$ of the linear ODE (\ref{vebpm}) the right side of the relation (\ref{rozh}) (relations
(\ref{unich})) satisfies ODE  (\ref{vebpm}) with the minus (plus) sign, and as $r_1\neq0$ ($r_2\neq0$) the identity (\ref{unich}) (the identity (\ref{rozh})) follows from the identity (\ref{rozh}) (respectively, (\ref{unich})).

Now the validity of the statement  formulated above Remark 3 is clearly seen after the change
$$\tau=it,\quad z=\sqrt{\frac{1}{\hbar}}x,
\quad \lambda=\sqrt{\frac{1}{\hbar}}q,\quad r_j=\sqrt{\frac{1}{\hbar}}c_j\quad (j=1,2),\quad A_\pm=Q_\pm.$$

\section { ``Quantum'' approach to the classification of Hamiltonian ODE integrated by the method of isomonodromic deformations}
Second-order ODE for the variable $\lambda$ following from the Hamiltonian system (\ref{Hams}) by excluding the impulse $\mu$  in case of the Hamiltonian (\ref{hamv}) is of the form
\begin{equation}\label{hodu}
\lambda''_{\tau\tau}=K(\tau, \lambda)(\lambda'_\tau)^2+L(\tau,\lambda)\lambda'_{\tau}+M(\tau,\lambda),
\end{equation}
where
$$K(\tau, \lambda)=
\frac{\alpha'_{\lambda}(\tau,\lambda)}{2\alpha(\tau,\lambda)},\quad
L(\tau,\lambda)= \frac{\alpha'_{\tau}(\tau,\lambda)}{\alpha(\tau,\lambda)}.$$
The change ($\lambda_*$ is a constat)
$$\varphi=\int_{\lambda_*}^{\lambda}\frac{d\nu}{\sqrt{\alpha(\tau,\nu)}}$$
transforms it into the equation of the form
\begin{equation}\label{normd}
\varphi''_{\tau\tau}=f(\tau,\varphi).
\end{equation}

The result of the linearization of ODE (\ref{hodu})
$$\Lambda''_{\tau\tau}=[2K(\tau,\lambda)\lambda'+L(\tau, \lambda)]\Lambda'_{\tau}
+[K'_{\lambda}(\tau,\lambda)(\lambda')^2+L'_{\lambda}(\tau,\lambda)\lambda'+M'_{\lambda}(\tau,\lambda)]\Lambda$$
in the general case depends not only on the coordinates $\lambda$, but also on the impulses $\mu$. It is natural therefore under the ``quantum'' linearization of ODE (\ref{hodu}) to associate to it a partial differential equation
$$W''_{\tau\tau}=A(\tau,z,\varphi)W_{zz}''+
D(\tau,z,\varphi)(W_z')'_{\tau}
+[E_1(\tau,z,\varphi)\varphi'+E_0(\tau,z,\varphi)]W'_{\tau}+$$
\begin{equation}\label{genklin}+[F_1(\tau,z,\varphi)\varphi'+F_0(\tau,z,\varphi)]W'_z+[J_2(\tau,z,\varphi)(\varphi')^2+
J_1(\tau,z,\varphi)\varphi'+
J_0(\tau,z,\varphi)
]W,\end{equation}
whose coefficients depend analytically on $\varphi$.

For the classification of Hamiltonian ODE (\ref{hodu})  integrated by the method of isomonodromic deformations the equation (\ref{genklin}) is suggested to be  a pair for the equation
\begin{equation}\label{kuangen}W'_{\tau}=\frac{W_{zz}''}{2}-[G(\tau,z)+R(\tau, \varphi,\varphi')]W
\end{equation}
(after simple substitutions the ``quantization'' (\ref{qquant}) of the ODE (\ref{hodu}), determined by the Hamiltonian (\ref{hamv}), is reduced to this equation) together with the
condition of compatibility of this pair on the solutions $\varphi(\tau)$ of the ODE (\ref{normd}). For making this classification the specific nature of the dependence of the equation (\ref{genklin}) on $\varphi'$ is essential, since the functions $\varphi$ and $\varphi'$ should be considered as independent variables. And the dependence of the function $R(\tau, \varphi,\varphi')$ on its arguments is not supposed to be analytical in advance (it is not excluded,  for instance, that this function can be described by means of nonlocal properties, i.e. by the integrals w.r.t. the variable $\tau$ of the combinations of the arguments).

It is obvious that the procedure of the ``quantum'' linearization of ODE being introduced is not strictly formalized. But the form of the equation (\ref{genklin}) looks  quite general and at the same time it reflects the nature of the described procedure.

Saying generally, for all the six Painlev\'e equations, in particular, the equations of the kind (\ref{genklin}) with $A=D=F_j=0$ i.e., linear ODE w.r.t. the variable $\tau$ are compatible with the equation (\ref{kuangen}). But

 1) for instance, the Painlev\'e equation IV  \begin{equation}\label{Pefo}\lambda''_{\tau\tau}=\frac{(\lambda'_{\tau})^2}{2\lambda}+ \frac{3\lambda^3}{2}+4\tau\lambda^2+2(\tau^2+4b)\lambda-\frac{8a+2}{\lambda}\end{equation}
($a,b$ are constants) and its Hamiltonian
\begin{equation} \label{Hamfo}H_{IV}(\tau,\lambda,\mu)=2\lambda \mu^2-\frac{\lambda^3}{8}-\frac{\tau\lambda^2}{2}-\frac{(\tau^2+4b)\lambda}{2}-2\frac{a+1/4}{\lambda}\end{equation}
alongside with the equation
\begin{equation}\label{khamfo}\Phi'_{\tau}=2z\Phi''_{zz}+2\Phi'_z-(\frac{z^3}8+\frac{\tau z^2}{2}+\frac{(\tau^2+4b)z}{2}-2\frac{a+1/4}{z}+H_{IV}(\tau,\lambda,\mu))\Phi,
\end{equation}
determined by ``quantization'' from \cite{Difufa},
are natural to be associated not with ODE but with the equation
\begin{equation}\label{qlinfo} \varepsilon^2\Phi''_{\tau\tau}=2\varepsilon  \Phi''_{z\tau}+\varepsilon\frac{\lambda'}{2\lambda}\Phi'_{\tau}-2\frac{\lambda'}{\lambda}\Phi'_z+
[\frac{z^2+3z\lambda+5\lambda^2}{2}+
\tau(6\lambda+2z)+2(\tau^2+4b)+
\frac{8a+2}{\lambda z}]\Phi(\tau)\end{equation}
as $\varepsilon=2$. The result of the linearization of the equation (\ref{Pefo})
$$\Lambda''=\frac{\lambda'}{\lambda}\Lambda'+[-\frac{(\lambda')^2}{2\lambda^2}+\frac{9\lambda^2}{2}+8\tau\lambda+2(\tau^2+4b)+\frac{8a+2}{\lambda^2}]\Lambda$$
appears exactly from the equation (\ref{qlinfo}) after formal change (the function $\mu=\lambda'/(4\lambda)$ is the classical impulse of the Hamiltonian (\ref{Hamfo}))
$$z\to \lambda,\quad\varepsilon^2\Phi''_{\tau\tau}\to \Lambda'',\quad \varepsilon(d\Phi/d\tau) \to \Lambda',\quad d/dz \to \mu=\lambda'/(4\lambda(\tau)),\quad \Phi \to \Lambda(\tau).$$
At the same time exactly this partial differential equation is satisfied by the simultaneous solution of the equation equation (\ref{khamfo}) and the equation $2(z-\lambda)\Phi_{\tau}=4z\Phi'_z-\lambda'\Phi.$
The latter pair of the equations is equivalent to the $L,A$ pair for the fourth Painlev\'e equation from \cite{Ufa88}, \cite{Difufa}.  The similar remark concerns Painlev\'e equation of the thirty-fourth type  connected with the second Painlev\'e equation, for its ``quantizations'' see \cite{Garfn};

 2) in the process of solving the problem of the classification it is supposed to list  also evolutionary equations (\ref{kuangen}), which for the solutions of ODE (\ref{normd}) (from the classical point of view, probably, as trivial as ODE (\ref{harmk})), are compatible with linear equation of the type (\ref{genklin}). For the solutions of evolutionary equations of the type (8) corresponding to different reductions of Painlev\'e equations see
[17], [28], [29].

Therefore it is not reasonable to restrict essentially in advance the form of equations (\ref{genklin}). However, it is possible  that while making the classification based on compatibility of solutions of ODE (\ref{normd}) of  $L,A$  pair (\ref{genklin}), (\ref{kuangen}), it happens to be necessary to impose also additional postulates, for instance, the validity of the identity $$W_z'\equiv[\varphi'\nu(\tau,\varphi) +\xi(\tau,\varphi)]W,$$
 on the curves $z=\varphi(\tau)$ which 
reflects the fact that for all the six Painlev\'e equations the solutions of their ``quantizations'' from the papers \cite{Ufa88} --- \cite{Garfn}  possess on such curves a property following from the validity  of  relations like second equation in (\ref{rashpeod}) for such solutions.

\section {Conclusion}

In conclusion we  note  that D.P. Novikov in his paper \cite{Novd} described relationships of  $L,A$   pairs for multi-component Hamiltonian systems of Schlesinger ODE \cite{Schl}  with some results of the papers \cite{Zam}, \cite{Fat}. But the natural problem on principal possibility of employing such relationships with some general ``quantum'' equations for the classification of multi-component Hamiltonian systems of ODE, admitting application of the method of isomonodromic deformation, is still to be considered.

The author is grateful to D.A. Polyakov for his favourable interest to the above results.

\end{document}